\documentclass[conference]{IEEEtran}
\ifCLASSINFOpdf
\else
\fi
\usepackage{url}
\usepackage{mathptmx}
\usepackage{amsmath}
\usepackage{mathptmx}

\usepackage{times}
\usepackage{xcolor}
\usepackage{lastpage}
\usepackage{amsmath}
\usepackage[utf8]{inputenc}
\usepackage{epsfig,endnotes,url,color}
\usepackage{array}
\usepackage{multirow}
\usepackage{xspace}
\usepackage{verbatim}
\usepackage{wrapfig}
\usepackage{mathptmx}
\usepackage[english]{babel}
\usepackage{graphicx}
\usepackage{csvsimple}
\usepackage{epstopdf}
\usepackage{subcaption}

\usepackage{enumitem}
\usepackage{datetime}
\usepackage{url}
\usepackage{cleveref}
\usepackage{amssymb}
\usepackage{tabularx}

\RequirePackage{snapshot}

\begin{document}
\title{Participation Cost Estimation: Private Versus Non-Private Study}

\author{
\IEEEauthorblockN{Joshua Joy\IEEEauthorrefmark{1},Sayali Rajwade\IEEEauthorrefmark{2},Mario Gerla\IEEEauthorrefmark{3}}
\IEEEauthorblockA{UCLA\\Email:\IEEEauthorrefmark{1}jjoy@cs.ucla.edu,\IEEEauthorrefmark{2}sayalirajwade@cs.ucla.edu,\IEEEauthorrefmark{3}gerla@cs.ucla.edu}
}


%Telephone: (800) 555--1212, Fax: (888) 555--1212}


\maketitle

% \fi
\IEEEpeerreviewmaketitle

\begin{abstract}

In our study, we seek to learn the real-time crowd levels at popular points of interests based on users continually sharing their location data. We evaluate the benefits of users sharing their location data privately and non-privately, and show that suitable privacy-preserving mechanisms provide incentives for user participation in a private study as compared to a non-private study.

\end{abstract}
\section{Introduction}

Today's smartphones not only serve as a means of communication and computation, though they also include a variety of sensors (e.g., proximity, accelerometer, gyroscopic, GPS, etc.).  Collection of this sensor information helps in the statistical analysis of the user's data. Various data analysts perform data aggregation and try to extract meaning from this data.

However, with this wide array of sensors comes serious privacy concerns with the large amount of personal real-time data that is collected without individual's consent or knowledge. Of the more serious privacy violations is with regards to location data. While location based services are rising in popularity, the concerns of being constantly tracked require proportional privacy-preserving mechanisms. There needs to be a mechanism by which the analyst should be able to receive the aggregated data in order to perform analysis but preserving the user's privacy at the same time. 

In this paper, we examine a study which collects real-time location data allowing individuals to see how crowded particular points of interest are. Our cost is a function of time in our study. The privacy mechanism we choose is that of differential privacy, which distorts the aggregate sum in such a way to hide the presence or absence of an individual.

Contemporary applications (e.g., Waze) continously collect and store location data without regards to preserving individual privacy. Our goal is to examine whether we can align the incentives of individuals wishing to  maximize their time efficiency with the goals of privacy. Participation in the non-private study allows individuals to optimize their time usage at the expense of being constantly tracked. Our evaluation demonstrates that the cost of the non-private study is higher than the private study. Thus, it is more beneficial for individuals to participate in the private study.

\section{Goals}

\begin{figure}[t!]
\centering
\includegraphics[scale=0.28]{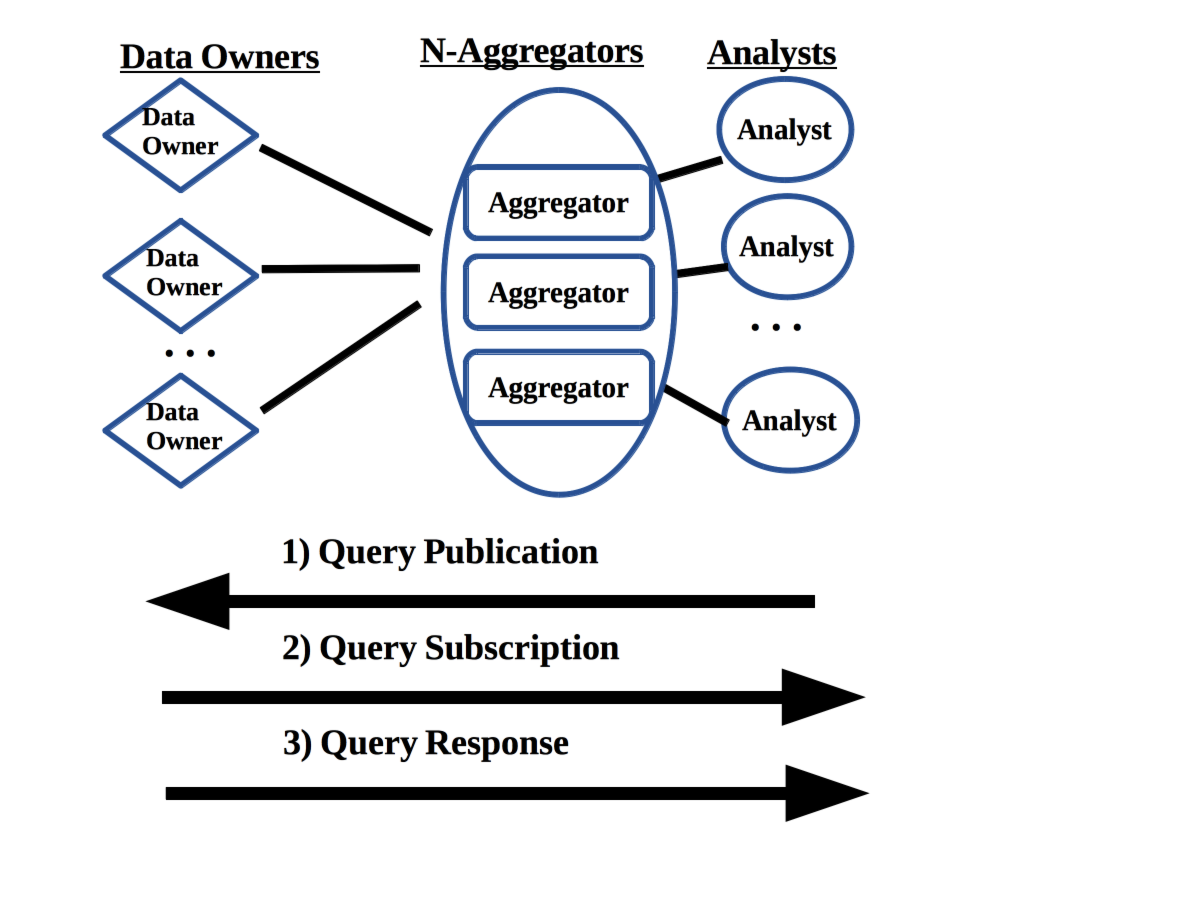}
\caption{System overview.}
\label{fig:systemoverview}
\end{figure}

We now describe the system goals, performance goals, threat model, and privacy goals of the architecture as seen in Figure~\ref{fig:systemoverview}.

\subsection{System Goals}

The system should support analysts who wish to run a population study. The analysts issue a query for those interested data owners that privately reply. Analysts are able to formulate long-standing signed queries. These queries continually elicit privatized responses during the defined query epoch. The analysts are deemed to be reputable, e.g., Department of Transportation, National Institutes of Health, or Centers for Disease Control. Each analyst controls an aggregation server.

We propose to conduct a study which will help individuals estimate the amount of time that will be spent at a particular location(e.g. gym, library, restaurant etc.) in and around campus based on the congestion (number of individuals present) at that location. In order to find the number of users present at a location, it is essential that the data owners publicly reveal their location information. A survey/study needs to be conducted which asks the data owners whether they were present at a given location at a specific time. In our study, it will be completely the data owner's choice whether or not to participate. Based on the number of individuals at each location, we will be able to tell for each location, how much time should an other individual wait before going to that location in order to save time and avoid rush. Only the individuals who decide to participate in the survey have access to this information. Here, the incentive for the users to participate in the survey is the time that they would be saving, if they know in advance the rush at a particular location.

The queries are propagated using roadside WiFi units or LTE repeaters. The requests may also be piggybacked on responses to information requested by the data owners, or in response to periodic polls. To reduce traffic overhead, or in response to periodic polls the queries may be posted to an edge website that mobiles of a certain class frequently check (e.g., campus  website).

The long-standing queries are needed to be fetched only once by each data owner. The responses of data owners and aggregation processing proceeds in epochs. That is every epoch each data owner privately and anonymously transmits their respective answer to the aggregator servers. The aggregation servers then compute the final aggregate using the received responses within this epoch. Epochs are defined on the order of seconds.

As there is a tradeoff between privacy and utility, the system should strike a superior balance the tradeoff of strong privacy and utility. We strive for a small percentage of error for both sparse and large datasets.

\subsection{Performance Goals}

The system should scale to hundreds of thousands of users and be able to process results on the order of seconds. Network bandwidth should be minimal as compared to existing bandwidth hungry systems \cite{DBLP:journals/joc/Chaum88,DBLP:conf/sosp/HooffLZZ15}.

\subsection{Threat Model}

Each aggregation server is owned by a set of distinct reputable analysts. Aggregation servers are expected to be available and online, so we do not consider denial of service attacks whereby data owners are not able to transmit their responses. We assume aggregation servers are honest-but-curious, i.e., servers do not corrupt the messages though can attempt to read all messages.

\subsection{Privacy Goals}

We assume all queries are signed and from reputable analysts. This provides provenance in the case of a dishonest analyst that may formulate a specially crafted query that attempts to deprivatize a data owner.

Data owners' privatized location responses should leak no more data than if they were not participating in the population study. Each data owner retains their own data on devices that they control and manage. The data owners then choose to participate in responding to each query. All responses before they leave the data owner are privatized. The privacy mechanism should satisfy the differential privacy criteria.

\section{Preliminaries}

\subsection{Differential Privacy}

Differential privacy has become the \emph{gold standard} privacy mechanism which ensures that the output of a sanitization mechanism does not violate the privacy of any individual inputs.  A privacy mechanism $San()$ provides $\epsilon$-differential privacy~\cite{DBLP:conf/icalp/Dwork06,DBLP:conf/tcc/DworkMNS06} if, for all datasets $D_1$ and $D_2$ differing on at most one record, and for all outputs $O \subseteq Range(San())$:
\begin{equation}
\Pr[San(D_1) \in O] \leq e^{\epsilon} \times \Pr[San(D_2) \in O]
\label{eqn:dp}
\end{equation}

That is, the probability that a privacy mechansim $San$ produces a given output is almost independent of the presence or absence of any individual record in the dataset.  In other words, it is difficult to determine whether any individual record is in the dataset, thus protecting privacy.  The privacy parameter $\epsilon$ controls the tradeoff between the accuracy of a privacy mechanism and the strength of its privacy guarantees: smaller $\epsilon$ provides stronger privacy but lower accuracy, and vice versa.

\begin{figure}[t!]
\centering
\includegraphics[scale=0.28]{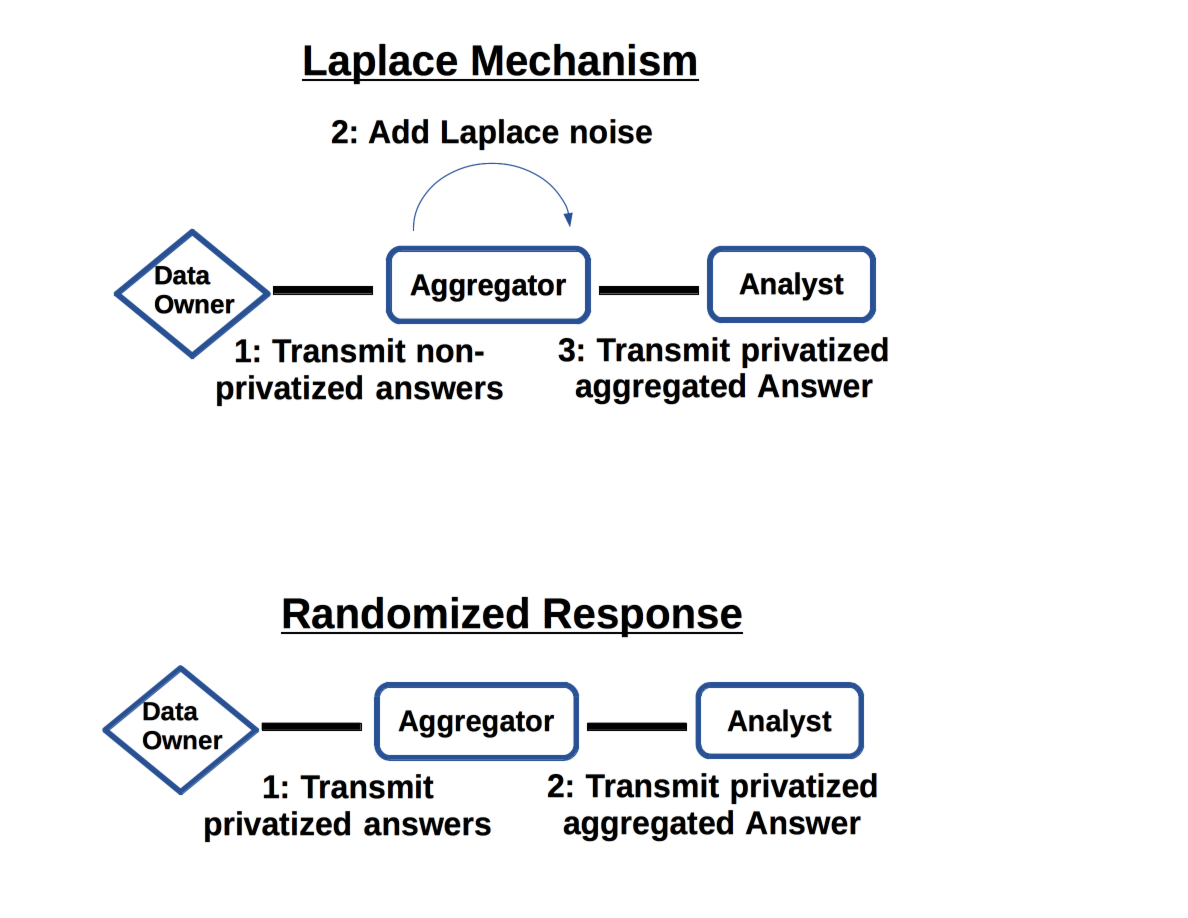}
\caption{Privacy mechanisms.}
\label{fig:mechanism}
\end{figure}

There are two differential private mechanisms we consider: the laplace mechanism and randomized response as seen in Figure~\ref{fig:mechanism}.

\subsubsection{Laplace Mechanism}

The Laplace privacy-preserving mechanism \cite{DBLP:conf/icalp/Dwork06,DBLP:conf/tcc/DworkMNS06,DBLP:journals/fttcs/DworkR14} is a centralized mechanism whereby an aggregator collects and stores the original, non-privatized data and then releases privatized answers to queries. The aggregator samples from the Laplace distribution such that the aggregated response $R$ is simply the sum of all the non-privatized responses $a$ and the Laplace noise $n$, $R = a + n$. The Laplace privacy-preserving mechanism was one of the earliest proposed differential privacy mechanisms, and provides the highest accuracy due to it's centralized nature.

%Dwork in [6] suggests the use of Laplace based noise addition in a query/response situation. Assume that the intruder issues a query f(X) on a data set X for which the true response is a. Let a differentially private mechanism Kf() be implemented for this data set and that the response from the system is R. Dwork in [6] suggests that the masked response R = a + y, where y represents a noise term from a Laplace distribution with mean 0 and scale parameter b = ?f/? where ?f represents the maximum difference in the value of f(X) when exactly one input to X is changed. This accounts for the situation, for example, when the intruder?s data differs from that of the data set X by exactly one record.

\subsubsection{Randomized Response}

Randomized response~\cite{warner1965randomized} was originally created by social scientists as a mechanism to perform a population study over sensitive attributes (such as drug use or certain ethical behaviors). Randomized response allows data owners to locally randomize their truthful answer to analyts' sensitive queries and respond only with the privatized (locally randomized) answer. Randomized response satisfies the differential privacy guarantee for individual data owners and  it provides the optimal sample complexity for local differential privacy mechanisms~\cite{DBLP:conf/nips/DuchiWJ13}.

So far, the randomized response mechanism has been widely adopted by both social scientists and computer scientists \cite{warner1965randomized,DBLP:conf/ccs/ErlingssonPK14}. There are many different randomized response mechanisms in the literature.  In this section, we present only the mechanism described in~\cite{fox1986randomized} because it strikes a superior balance between the utility and the privacy guarantee of randomized responses, as compared to other mechanisms~\cite{warner1965randomized,kuk1990asking,greenberg1969unrelated,greenberg1971application}.

\textbf{Mechanism Description} Suppose each data owner has two independently biased coins. Let the first coin flip heads with probability $p$, and the second coin flip heads with probability $q$.  Without loss of generality, in this paper, heads is represented as ``yes'' (i.e., 1), and tails is represented as ``no'' (i.e., 0).

Each data owner flips the first coin. If it comes up heads, the data owner responds truthfully; otherwise, the data owner flips the second coin and reports the result of this second coin flip.

Suppose there are $N$ data owners participating in the population study. Let $\hat{Y}$ represent the total aggregate of ``yes`` randomized answers. The estimated population with the sensitive attribute $Y_A$ can be computed as:
\begin{equation}
\label{eqn:yo}
Y_A = \frac{\hat{Y} - (1 - p) \times q \times N}{p}
\end{equation}

The intuition behind randomized response is that it provides ``plausible deniability'', i.e., any truthful answer can produce a response either ``yes'' or ``no'', and data owners retain strong deniability for any answers they respond. If the first coin always comes up heads, there is high utility yet no privacy. Conversely, if the first coin is always tails, there is low utility though strong privacy. As we show in in Table~\ref{tab:utility-privacy}, by carefully controlling the bias of the two coin flips, one can strike a balance between utility and privacy.

\begin{table}[t!]
\small
\centering
\begin{tabular}{c|c|c|c}
\hline $p$ & $q$ & Relative Error ($\eta$) & Privacy Level ($\epsilon$) \\
\hline
\hline \multirow{3}{*}{0.3} & 0.3 & 0.1958 & 0.8873\\
  & 0.6 & 0.1833 & 0.5390\\
  & 0.9 & 0.1333 & 0.3895\\
\hline \multirow{3}{*}{0.6} & 0.3 & 0.0958 & 1.7918\\
  & 0.6 & 0.0875 & 1.2528\\
  & 0.9 & 0.0708 & 0.9808\\
\hline \multirow{3}{*}{0.9} & 0.3 & 0.0569 & 3.4340 \\
  & 0.6 & 0.0542 & 2.7726\\
  & 0.9 & 0.0514 & 2.3979\\
\hline
\end{tabular}
\vspace{-2mm}
\caption{The utility and the privacy guarantee of query results with different coin flipping probabilities $p$ and $q$.  Here, utility is measured by the query result's relative error, and privacy is measured by the level of achieved differential privacy.  Before randomization, there are 1000 original answers, 80\% of which are ``yes''.  Results are generated based on 100 independent runs.}
\vspace{-2.5mm}
\label{tab:utility-privacy}
\end{table}

\section{Evaluation}

\begin{figure}[t!]
\centering
\includegraphics[scale=0.45]{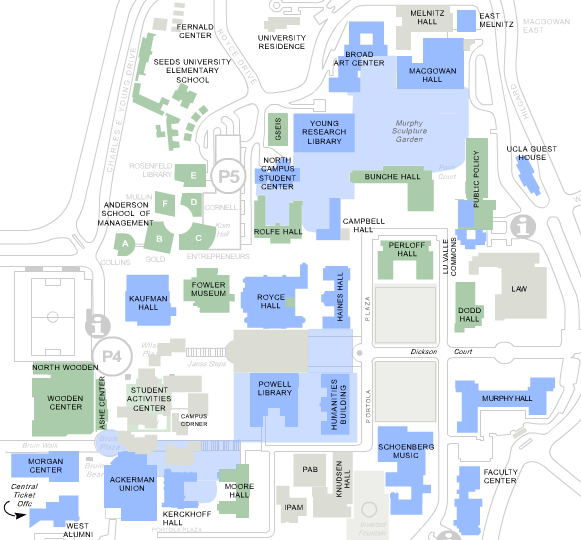}
\caption{Map.}
\label{fig:map}
\end{figure}

Our case study evaluates the campus scenario as seen in Figure~\ref{fig:map} whereby the points of interests include gyms, cafeterias, and libraries.  A differentially private study is conducted to measure real-time congestion levels whereby the collected data is immediately released to the public.

Let us now consider the scenarios in which the individual is offered a choice between participating in the study and declining, however the study will always take place. The campus under study has an estimated 30,000 undergraduates and 12,000 graduates. We estimate that around 5,000 out of the total 42,000 students decide to participate in the study and around 2,000 decide to not participate. In order to calculate the costs, as measured by time, in each of the individual scenarios, we need to calculate the base cost and the worst case cost that the individuals might incur. Lets call $W$, the worst case cost and $E$, the base cost. 
	The worst case cost $W$ would be the maximum amount of time that they would have to spend at a particular location during the peak hours. Let us assume that cost $W$ in terms of number of minutes. Let $W = 60$,  i.e. the individual has to wait at the location for a maximum of one hour at the peak hour. If the individuals decide to not participate in the study, then the probability of getting penalized by congestion at a particular location would be high, because they would not be using the results about congestion provided by the study. The study might have some error estimating the congestion at a location, let us assume that this probability $p$ is $0.5$. 
	The base cost, if the participants decide to not participate in the study would be the probability considered above, times the worst case cost $W$.

\begin{equation}
E = p * W
   =  0.5 * 60
   = 30
\label{equn:worstcase}
\end{equation}

We can now calculate the cost for both the scenarios individually.

\subsection{Case 1: Individual does not participate}

If the individual decides to not participate in the study, then they will become penalized by the congestion at the location. Also, since these individuals are not a part of the private study, they have a higher probability of their location information being disclosed. Let $\phi$ be the fraction of the participants who might be de-anonymized. Let us consider this value to be 0.8 (i.e. 80 out of 100 participants are re-identified). Therefore, the cost function when the individual decides to not participate in the study (or we can also call this as, the cost function of a non-private study) can be given by the following equation described in \cite{DBLP:conf/csfw/HsuGHKNPR14}:

\begin{equation}
C = \phi * W * N
\end{equation}

whereby

$N$ is the number of participants who decide to not privately participate. Let $N = 2000$ as mentioned above.

Therefore plugging in the values of $\phi$, $W$ and $N$ in \cite{DBLP:conf/csfw/HsuGHKNPR14}, the cost would be:

\begin{equation}
C = 0.8 * 60 * 2000 = 96000
\end{equation}

\subsection{Case 2: Individual participates in the study}

If the individual decides to participate in the study then the base case cost would be reduced by some factor of $\epsilon$, which is the privacy level in a differentially private study.  Therefore the overall cost function as mentioned in \cite{DBLP:conf/csfw/HsuGHKNPR14} for a private study can be given by the following equation:

\begin{equation}
C  = (e^\epsilon - 1) * E * N
\end{equation}

whereby

$N$ is the number of participants who decide to participate in the study. Let $N = 5000$ as mentioned above. 
$E = 30$ as given in equation \cite{DBLP:conf/csfw/HsuGHKNPR14}.

\subsection{Choosing Epsilon}

In our study, the individuals respond to the query with randomized responses, using two coin flips/tosses. The value of $\epsilon$ depends on the coin bias probabilities and the utility which is the relative error. Table~\ref{tab:utility-privacy} shows different values of epsilon for different coin bias probabilities ($p$ for the first coin and $q$ for the second coin) and relative errors.

Plugging in the values of $E$, $N$ and a few different values of $\epsilon$, we can obtain various values of cost. These values are shown Table~\ref{tab:costepsilon}.

As we can see from the table, the cost decreases as we go on reducing the value of $\epsilon$. We need to compare these values of cost with the cost obtained in the first scenario where the individual decides to not participate in the study. The cost of our private study needs to be less than the cost of the non-private study i.e. the cost when the individual decides to not participate in the study. A lower cost indicates less amount of time spent at the congested locations.

The value of cost will help us determine the value of epsilon for our study. From the table above, we can say that for $\epsilon = 0.3895$, we obtain a cost of $71436$ which is less than the cost $96000$. Thus, the value of $\epsilon = 0.3895$ would be appropriate in our case.
	
\begin{figure}[t!]
\centering
\includegraphics[scale=0.45]{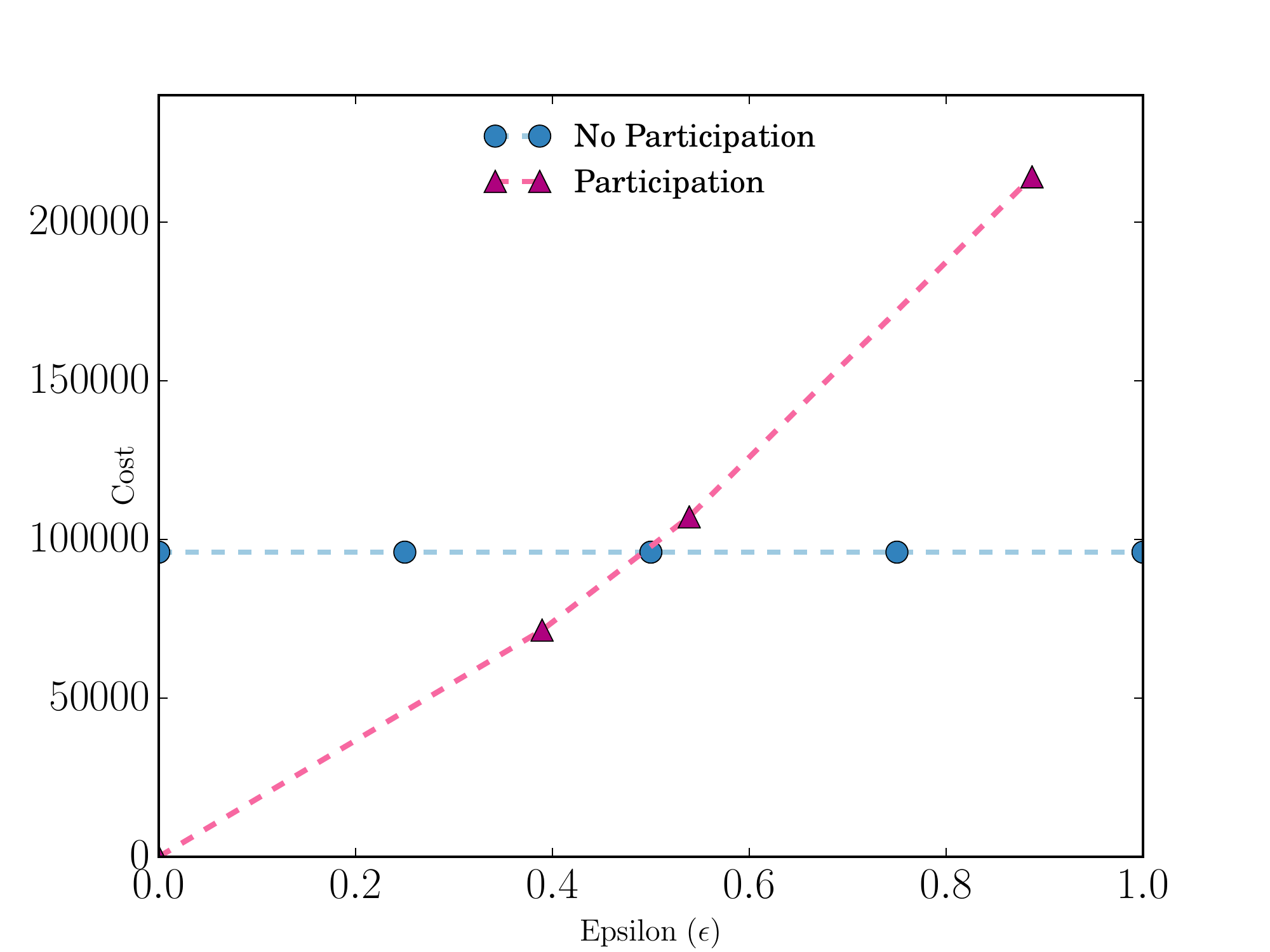}
\caption{Cost vs Epsilon.}
\label{fig:costvsepsilon}
\end{figure}

In order to analyze this further, we can plot the values of epsilon vs cost and compare it with the cost as seen in Figure~\ref{fig:costvsepsilon}. The figure shows that adjusting the privacy parameter $\epsilon$ influences the cost of participation. This is largely due to the inherent trade-off with privacy-preserving mechanisms of privacy versus accuracy. Greater privacy comes with a loss of accuracy and vice-versa.  The incentive for the individuals to participate in the private study is the lower value of cost, which is the amount of time they would spend at the congested location. From the plot, we can say that all the values of $\epsilon$ which fall below the non-participation cost would be ideal values. Thus in this case, the plot shows that all values below approximately $0.5$ are ideal values of $\epsilon$. 

The smaller the value of $\epsilon$, the better incentives there are for an individual to participate. For example, the $\epsilon$ value of $0.0083$ would have a cost of only $1250$ which is favorable for the individual. It's important to observe that randomized response is not able to achieve such small values of $\epsilon$. However, centralized aggregation mechanisms such as Laplace noise are able to achieve such small values. Thus, considering cost as defined here, we might want to say that the cost is better when the Laplace mechanism is used as compared to the randomized response. 

\subsection{Discussion}

The cost estimations are done according to very specific conditions in this study. Several different scenarios can be considered  where the congestion is going to vary based on different times of the day or different seasons of the year. The cost calculations depend on a number of factors such as $N$ the number of individuals participating and $E$ which is the base cost. These quantities might be very different for the different scenarios mentioned above. 
	
The differentially private study proposed in this report suggests the use of randomized response for adding differentially private noise to the output of the computation. However, as observed in the comparison between the Laplace mechanism and randomized response, the Laplace mechanism performs better. Yet, it is important to note the parameters of the cost function. There is a fundamental privacy notion difference between Laplace noise and randomized response in that the former is distributed and maintain local privacy while the latter is centralized and requires strong trust assumptions of the aggregator. It would be beneficial to consider factoring in the trust assumptions into the cost computation.

\begin{table}[]
\small
\centering
\begin{tabular}{l|l}
\textbf{Epsilon ($\epsilon$)} & \textbf{Cost (C)} \\
\hline
2.3979               & 1500008           \\ \hline
1.2528               & 375020            \\ \hline
0.8873               & 214285            \\ \hline
0.539                & 107144            \\ \hline
0.3895               & 71436            
\end{tabular}
\caption{Cost as a function of Epsilon ($\epsilon$).}
\label{tab:costepsilon}
\end{table}

\section{Conclusion}

We have examined a population study whereby individuals are able to choose to privately participate or not. The utility function for the individual is a cost function which factors the amount of time an individual may or may not spend at a particular congested location. We have calculated costs for two scenarios - one when the user decides to not participate in the study and the other when the user participates in the study. We demonstrate that by varying the privacy level $\epsilon$ we can achieve a lower cost (i.e., less time spent at the congested location) as compared to not participating in the privacy study. Additionally, we raise the question of factoring trust assumptions of services into the cost function as future work.

\bibliographystyle{acm}
\bibliography{fss,privacy,main,vehicles,mpc,cost}

\end{document}